# Space Weathering of the 3-µm Phyllosilicate Feature induced by Pulsed Laser Irradiation

B.S. Prince[1], and M.J. Loeffler[2,3]

[1]Department of Applied Physics and Materials Science, Northern Arizona University, Flagstaff, AZ, 86011

[2]Department of Astronomy and Planetary Science, Northern Arizona University, Flagstaff, AZ, 86011

[3]Center for Materials Interfaces in Research and Applications, Northern Arizona University, Flagstaff, AZ, 86011

## Abstract

Here we present results from pulsed laser irradiation of CI and CM simulant samples in an effort to simulate space weathering on airless bodies via micrometeorite impacts. For this study, we focused on determining what type of alteration occurs in the 3-µm absorption region, as this region will be critical to ascertain compositional information of the surface regolith of hydrated asteroids. Generally, using entirely *in situ* spectral analysis, we find that the laser produces similar effects in both samples. Specifically, irradiation causes the blue spectral slope to decrease until it is relatively flat and that the sample darkens initially with laser irradiation but brightens back to about half of its original level by the end of the irradiation. Furthermore, we also find that laser irradiation causes the band depth on the 3-µm absorption band to increase by as much as 30%, yet the shape of the entire absorption band does not change and the band minima of the 2.72 µm shifts less than 0.001 µm after laser irradiation. The constancy of the latter two parameters, which will be most critical to compositional analysis, suggests that this spectral region could be very useful to determine the asteroid composition on surfaces on hydrated asteroids that have undergone extensive aqueous alteration even if the surface had been subject to a significant amount of space weathering. Whether the same conclusion will be generally applicable to other surfaces containing minerals with a wide range of aqueous alteration is currently unclear but will be tested in future studies.

## 1. Introduction

Asteroid taxonomy has been a topic of study for decades (Gaffey et al. 1993, Bus and Binzel 2002, Demeo et al. 2009). Through use of primarily ground-based spectroscopy, the near-infrared spectral region (~0.5 and 2.5 μm) has been used to compare and classify asteroids based on their spectral characteristics (e.g., slope, albedo, and absorption bands). Gaining additional insight about asteroid surfaces via analysis of a different spectral region with ground-based spectroscopy is challenging, as telluric absorption features at slightly longer wavelengths can prevent any light that may have been reflected from the asteroid from reaching the telescope. Even with these difficulties, ground-based spectroscopy was used to make the first detection of a hydrated asteroid (Lebofsky 1978). More recent measurements also suggest that numerous asteroids may also contain hydrated minerals (Takir et al. 2017, Rivkin et al. 2019, Usui et al. 2019).

Being able to characterize the spectral reflectance of an asteroid surface at wavelengths longer than those typically studied could be particularly useful for characterizing the composition of carbonaceous asteroids, which are relatively dark and lacking in diagnostic features in the near-infrared region but are thought to be hydrated to various degrees.  The potential utility of infrared analysis for asteroid surfaces has been shown in analysis of carbonaceous chondrite samples that show a strong O-H stretch feature at 2.7 μm (Miyamoto and Zolensky 1994, Takir et al. 2013) that varies in size and shape, as well as in recent spacecraft measurements of Bennu (Hamilton et al. 2019, Barucci et al. 2020, Merlin et al. 2021, Praet et al. 2021, Zou et al. 2021) and Ryugu (Kitazato et al. 2019). It seems likely that unobstructed observation of this spectral region will become more common after the launch of James Webb Space Telescope, which should easily provide coverage of this spectral region (Milam et al. 2016, Rivkin et al. 2016).

One issue that complicates the analysis of any airless body's surface via remote sensing spectroscopy is that the surface of these airless bodies are likely altered through a variety of processes, including  irradiation by solar wind ions and fast interplanetary dust particle (micrometeorite) impacts. Together, these processes are referred to as space weathering (Clark et al. 2002, Chapman 2004, Gaffey 2010), as they can cause the surface of the body to undergo chemical and topographical changes, which, in turn, can alter the spectral properties of the asteroid. While these changes have typically been detailed in the visible and near-infrared region (Hapke 2001, Sasaki et al. 2001, Brunetto and Strazzulla 2005, Loeffler et al. 2009), it seems likely that there will be specific alterations associated with other spectral regions as well. Thus, understanding to what degree the hydrated absorption feature is altered by space weathering processing will be critically important as future efforts to characterize asteroids in this spectral region become more common.

Laboratory studies investigating space weathering at wavelength longer than 2.5 µm have been somewhat limited and have primarily focused on solar wind implantation into anhydrous minerals relevant to the lunar surface (e.g., (Zeller et al. 1966, Mattern et al. 1976, Burke et al. 2011, Ichimura et al. 2012, Schaible and Baragiola 2014, McLain et al. 2021). The studies that have focused on the hydrated features intrinsic to the mineral have shown that space weathering processing may alter this spectral region. For instance, pulsed laser irradiation of the Murchison meteorite showed that the hydrated absorption feature severely attenuates as a result of laser irradiation (Matsuoka et al. 2015, Matsuoka et al. 2020, Thompson et al. 2020). Additionally, a recent study of He$^+$ irradiation of various carbonaceous chondrite samples showed that the sharp 2.7 µm absorption feature attenuated significantly and the main band minimum shifted to slightly longer wavelengths as a result of ion irradiation (Lantz et al. 2017). A spectral shift in the band minimum was also suspected in earlier ion irradiation experiments of Murchison, although the presence of adsorbed $H_2O$ made estimates somewhat difficult (Lantz et al. 2015). Recently, ion irradiation experiments on Mg-rich serpentine found that although the shape and intensity of this absorption feature did not change significantly, there was slight but measurable shift of 0.015 µm in the band minimum of the 2.7 µm band (Rubino et al. 2020). Most recently, Nakauchi et al. (2021) showed that implantation of hydrogen into saponite and serpentine caused the absorption strength of the hydrated features near 3-µm to increase slightly with irradiation but no spectral shift in their band minimum was noted.

Thus, given the potential importance of the hydrated absorption feature for future characterization of asteroid surfaces and the few existing laboratory experiments on this topic, we performed laser irradiation of carbonaceous chondrite simulant samples, while monitoring the sample with *in situ* infrared spectroscopy and playing particular attention to changes in the absorption features peak position, shape and band depth. To our knowledge, this is the first study on the spectral region where laser irradiation and analysis are performed *in situ*. This should simplify analysis, as the contribution of terrestrial $H_2O$ to the spectrum, either via direct adsorption or possibly due to any reaction with the altered surface will be minimized.

## 2. Materials and Methods

*2.1 Sample preparation*

All experiments were performed on carbonaceous chondrite (CI and CM) pellet simulants. The simulants were obtained from Deep Space Industries and have been designed to replicate the mineralogical and spectral properties of various carbonaceous chondrites (Britt et al. 2019). To prepare our pellet samples, we dry-sieved simulant powder to grain sizes <45 µm and pressed the sieved powder into a pellet using a 12 mm aluminum retaining ring at a pressure of ~1500 psi for ~5 minutes. We opted to use the smaller grain size, compared to the 45 – 125 µm size fraction we use for laser irradiation (Loeffler et al. 2016, Prince et al. 2020), because prior characterization of these simulant grains shows that material larger than 50 µm are typically agglomerates of smaller grains (Britt et al. 2019). We chose to press the pellets at 1500 psi because visual inspection showed that they held together well during laser irradiation, while the ones pressed at lower pressures (e.g. 1000 psi) tended to disintegrate or contain visible ruts in the surface after a few laser pulses. The resulting pellet was about 10 mm in diameter and 2-3 mm thick.  The sample pellets were then mounted onto an aluminum holder, inserted into an oil-free ion-pumped chamber, and brought to a pressure of ~1 x $10^{-8}$ Torr (Prince et al. 2020).

*2.2 Irradiation and Analysis*

Laser irradiation is typically employed to replicate the heating that occurs during a micrometeorite impact and has been previously used by our group (Loeffler et al. 2008a, Loeffler et al. 2008b, Loeffler et al. 2016, Thompson et al. 2019, Prince et al. 2020, Thompson et al. 2020), as well as many others (Sasaki et al. 2001, Brunetto et al. 2006, Gillis-Davis et al. 2017).  A more detailed discussion on the relation between laser pulses and micrometeorite impacts may be found in our previous work (Loeffler et al. 2008b, Loeffler et al. 2016). In the studies presented here, laser irradiations were performed using a Continuum Minilite Nd-YAG laser, which emits 6-8 ns pulses (λ = 1064 nm) at 1Hz.  The laser was aimed at an incident angle of 45° with respect to the surface normal of the pellet sample, and the resulting elliptical irradiated spot was measured to have an area of ~0.6 mm$^2$, which corresponds to an energy fluence of 0.6 J cm$^{-2}$ per laser pulse at the chosen laser energy (3.5 mJ). The laser was rastered across the sample using an actuated mirror that was synchronized to the laser. In the following, one pulse corresponds to one sweep across the entire sample; each sweep across the sample took approximately two minutes.

IR reflectance analysis of the samples were conducted *in situ* before, after, and intermittently during laser irradiation, using the same experimental approach described in our previous work (Prince et al. 2020). However, to enable the analysis of the mid-infrared region, we modified the setup slightly, installing KBr windows on the vacuum chamber, replacing our teflon reference standard with a diffuse gold reference (Infragold from Labsphere), and changing the components in our Thermo-Nicolet IS50 Fourier Transform spectrometer to an infrared light source, an XT-KBr beamsplitter, and an MCT/A detector. Reflectance spectra are given by:

$$R = \frac{I_{sample}}{I_{reference}}$$

where $I_{sample}$ and $I_{reference}$ are the reflected intensity of the chondrite simulant pellet and the gold reference, respectively. Each raw spectrum is an average of 1600 scans taken at a resolution of 16 cm[-1]. For more details on our experimental setup and procedures see Prince et al., (2020).

Using the derived reflectance spectra, we can quantify some key spectral changes induced during laser irradiation of our samples. To assess general alteration of the sample spectra, we analyzed the overall albedo and spectral slope at 1.8 μm. We chose evaluate these two parameters at this wavelength, because there were no overlapping absorption bands or structure nearby, this region showed clearly quantifiable slope changes and this wavelength could also be easily observed by other studies focusing on the near-infrared region. The spectral slope reported in this study has been calculated by evaluating the derivative of the spectrum. The overall albedo has been calculated by comparing the initial reflectance of the sample ($R_0$) with the reflectance of the sample after laser irradiation ($R$). As the sample could potentially brighten or darken, we refer to this as the normalized reflected intensity, which is defined as $R/R_0 - 1$. Finally, and perhaps most importantly, we also quantify changes to the 3-μm absorption band by evaluating any shifts in the band minima and changes in the band depth during laser irradiation. The band position was quantified by both visual inspection of a continuum removed spectra and by taking the derivative of the spectra. The band depth is defined as $1 - R_B/R_C$ (Clark and Roush 1984), where $R_B$ is the reflectance at the band minimum and $R_C$ is the reflectance of the continuum at the band center. In both cases, a line was sufficient to properly reproduce the continuum beneath the 3-μm absorption band.

## 3. Results

### 3.1 Initial Experiments

One potential complication analyzing the 3-μm spectral region in our mineral samples is that terrestrial $H_2O$ may be incorporated into the structure of the mineral and/or adsorbed onto the mineral surface. To test the contribution from terrestrial $H_2O$ in our laboratory setup, we took the reflectance spectrum of our CI and CM samples under ambient conditions and compared those spectra to spectra taken after the samples were under vacuum (Figure 1 (top)). Although the spectra are qualitatively similar, the terrestrial $H_2O$ initially present in the sample causes the ambient spectra to have a slightly broader absorption at 3 µm and a deeper absorption band at 6 µm. This difference can be seen more clearly by subtracting the spectrum taken under vacuum by one taken under ambient conditions (Figure 1 (bottom)). We note that spectra taken immediately after evacuating our system very closely resembled the vacuum spectra shown in Figure 1 (top) and additional heating of our sample to ~150° C failed to change the spectrum further, suggesting that in our experimental setup UHV conditions alone are sufficient to remove the majority of the absorbed water from our samples. Our spectra taken under ultra-high vacuum for both the CI and CM samples are also consistent with the spectra reported by Britt et al. (2019), as they are dominated by a strong feature centered at 2.72 µm, which is attributed to O-H stretching in the various phyllosilicates that comprise the majority of each of the simulants. The measured slope of our samples is slightly more blue (negative) above 2-µm than those reported by Britt et al. (2019), which we attribute to a pressing effect in our pellet samples.

*3.2 Pulsed Laser Irradiation*

After characterizing our fresh pellet samples under vacuum, we irradiated them with a pulsed laser, while monitoring them with *in situ* reflectance spectroscopy (Figure 2). We find similar trends for both simulant samples, which is not unexpected, given their similar spectral characteristics in this region. Specifically, we find that after 1 pulse of irradiation, the overall albedo decreases, and this effect is maximized after ~2 pulses. Further irradiation caused the samples to brighten slightly towards their original level. In addition, laser irradiation also causes the blue spectral slope observed at lower wavelengths to decrease (become more red) with progressive laser irradiation, until it appears relatively flat. Focusing on the 3-µm absorption band, we find that the overall strength of the band appears to increase with laser irradiation, as do the other minor features at longer wavelength, yet the absorption minimum and general shape does not appear change significantly in either of the simulant samples.

To better quantify the spectral changes described above, Figure 3 shows the normalized reflected intensity at 1.8 µm, the spectral slope at 1.8 µm, and the band depth at 2.72 µm for both CI and CM samples as a function of the number of laser pulses. Both samples darkened by ~15% (Figure 3 top) during laser irradiation and the vast majority of this change occurred after a single laser pulse. The subsequent brightening was a reproducible effect and typically brought normalized reflected intensity about half-way up to its original level. The spectral slope increases (becomes more "red") with laser irradiation and until it is nearly flat, again, most of the slope change occurs after the first laser pulse. The band depth increases with laser irradiation in both samples and up to as much as 30% by the last laser pulse.

In addition to the band depth, we also examined whether the band center and the overall shape of the 3-μm absorption band were altered by laser irradiation. Evaluation of the derivative of the continuum removed spectra allows us to estimate that the sharp drop, as well as the band center are constant to within 0.001 μm at any point during our laser irradiation. We also find that there is minimal change to the overall shape of the entire absorption band, even though the band depth increases with laser irradiation (see above). This latter observation is better illustrated by comparing the continuum removed reflectance spectra for both CI and CM samples during laser irradiation, where it is clear that applying a simple scaling factor to the laser irradiated spectra allows us to essentially match the initial reflectance spectrum of either sample (Figure 4).

## 4. Discussion

*4.1 Complicating factors for Proper Analysis of the 3-μm absorption band*

While there have been relatively few space weathering studies focusing on the 3-μm absorption band in hydrated minerals samples, spectroscopic studies to perform basic characterization of this spectral region data back decades (Salisbury et al. 1991, Miyamoto and Zolensky 1994). The different samples and different experimental setups used to characterize the 3-μm absorption region have illustrated that even before alteration from simulated space weathering processing, there are a number of factors that can complicate analysis and should be considered before discussing possibly weathering-induced effects.

The most basic factor is the role of terrestrial $H_2O$, as multiple early studies were only able to acquire spectra under ambient conditions (Salisbury et al. 1991, Miyamoto and Zolensky 1994). More recent measurements that could measure the reflectance spectra of samples both under ambient as well as under vacuum conditions show that there are cases where terrestrial $H_2O$ may only make a minor contribution (e.g., Figure 1) but there are also cases where it could dominate the spectrum (Takir et al. 2013, Britt et al. 2019). In fact, measurements of the CI simulant under ambient conditions in our laboratory compared with Britt et al. (2019) shows very different spectra, suggesting that the degree to which in situ analysis would be required to produce a representative starting spectrum will be strongly dependent on the local environment (humidity, etc.). This difference is likely attributable to the smectite-group clays present in the CI simulant, as it has been shown that the quantity of interlayer water in clays is a function of relative humidity (Moore and Reynolds Jr. 1989). These higher humidity conditions will also cause weaker absorption bands of $H_2O$, such as the fundamental bending mode at 6-μm (see

Figure 1) to become more prominent, which could potentially complicate analysis in other spectral regions as well.

Another factor that has been clearly demonstrated is that the degree of aqueous alteration can significantly change both the shape and position of 3-μm absorption band (Beck et al. 2010, Takir et al. 2013, Takir et al. 2019). Specifically, the absorption band narrows and moves to shorter wavelengths as aqueous alteration increases, which is what is observed in Mg-rich serpentines, such as our simulant samples or other CI meteorites (Beck et al. 2010, Takir et al. 2013). The least aqueously altered minerals (e.g., Fe-rich serpentines) show a broader absorption band than do the Mg-rich serpentines located near ~2.8 μm. Although the band minimum is still at significantly shorter wavelengths than terrestrial $H_2O$ (3.0 μm), such that the two forms of hydration should be able to be distinguished, the broad nature of both absorptions make determining whether or to what degree terrestrial $H_2O$ contributes to these spectra harder to quantify if spectra are measured under ambient conditions.

Finally, it is also important to consider that even within a single meteorite sample heterogeneity exists, which could potentially manifest itself in this spectral region. While multiple measurements for the same meteorite sample are limited, Murchison has already been noted to exhibit spectral heterogeneity in the near-infrared (Cloutis et al. 2011). Furthermore, reported spectra in the mid-IR show that the band shape of the 3-μm absorption features varies significantly (Beck et al. 2010, Lantz et al. 2015, Matsuoka et al. 2015, Takir et al. 2019). While some of this difference may be due to terrestrial $H_2O$ (see below), even within the same setup (Lantz et al. 2015), two separate Murchison samples showed considerably different spectral characteristics. Whether this was due to variations in the ambient environment, sample surface conditions or a real consequence of heterogeneity could be tested in future studies.

*4.2 Similarity of the CM and CI Simulants*

As shown in Figure 2, the initial spectra of the CI and CM simulants are very similar, with the CI simulant being slightly darker across all wavelengths, which is likely a consequence of the higher wt% of dark material present in the CI sample. Specifically, there is more pyrite (6.5 vs. 2.6 wt%), coal (5.0 vs. 3.6 wt%) and magnetite (13.5 vs. 10.4 wt%) in the CI simulant compared with the CM simulant (Britt et al. 2019). Other potentially key differences in the two samples is that the CI simulant contains vermiculite (9 wt%) and epsomite (6 wt%), while the CM simulant does not (Britt et al. 2019), However, the observation that laser irradiation produces similar spectral changes in the CI and CM simulant samples (Figure 2 and 3) suggests that these minerals either  do not play an important role in space weathering in this spectral region or are at concentrations where the spectral changes do not depend strongly on composition.

Besides the compositional differences noted above, the main contributor to the 3-μm absorption region and the dominant mineral in both simulant samples is serpentine. However, as Fe-rich serpentines, which would have ideally been used in the CM simulant, are not easily found in terrestrial minerals, the same type of Mg-rich serpentine was used for simulant sample (Britt et al. 2019). Thus, our observation that the 3-μm absorption region looks nearly identical and responds similarly to laser irradiation should not be surprising. Furthermore, it is also suggests that at least from the perspective of the 3-μm region, the CM simulant sample is more representative of aqueously altered CI meteorites than the CM group.

### 4.3 Comparisons to Previous Space Weathering Experiments

Laboratory studies investigating how space weathering processes may alter the 3-μm absorption band in hydrated minerals have been relatively limited. Below we compare our results to these studies, focusing on weathering trends that may be of interest for remote sensing but keeping in mind the difficulties and potential complications associated with this work.

### 4.3.1 Previous Laser Irradiation Experiments

The only studies that we are aware of using pulsed laser irradiation and performing analysis in the 3-μm were performed on Murchison meteorite chips (Thompson et al. 2020) and powders (Matsuoka et al. 2015, Matsuoka et al. 2020). Given the different scattering properties of these two types of samples, we will mainly focus our discussion on the previous powder experiments, as they are more directly comparable to our studies. Analysis of the irradiated Murchison powders were performed in a purged $N_2$ environment while heating the sample to 100 $^0$C (Matsuoka et al. 2015, Matsuoka et al. 2020) and showed that laser irradiation caused significant darkening of Murchison's reflectance spectrum between 0.5 – 8 μm, as well as severe attenuation of the 3-μm absorption band. The initial darkening, although more significant in the Murchison sample, is consistent with our results and is a trend that has been commonly observed at shorter wavelengths for laser irradiated of silicate minerals (Yamada et al. 1999, Sasaki et al. 2001, Brunetto et al. 2006, Loeffler et al. 2016). Matsuoka et al. (2015) attributed darkening of Murchison to FeS-rich amorphous silicate particles that have coated the surface either through splash melting or vapor deposition. Given that direct laser irradiation of FeS, as well as vapor redeposition of pure FeS onto FeS actually increases the reflectance of this mineral (Prince et al. 2020), we suspect that another mechanism may be important. One possibility is that the olivine and magnetite present in our simulant samples could be producing chemically reduced nanophase iron that could darken the spectra (Hapke 2001). Another possibility, recently noted by Matsuoka et al. (2020), is that structural (amorphization) and chemical (dehydration) changes induced by the laser within serpentine may also lower the albedo. Even though we do not see any indication of dehydration in our simulant samples, it seems likely that the laser will produce an amorphous melt, which could potentially lower the albedo as well (Matsuoka et al. 2020). Finally, in addition to the initial darkening, we also observe that after 5 laser pulses both samples brighten about half-way to their original level. The origin of this brightening is unclear from spectroscopy alone, but we note that a similar observation has been attributed compositional changes in the organics present in asphaltite (Moroz et al. 2004) and possibly Murchison (Thompson et al. 2020), suggesting something similar may be happening in our carbon (coal) component. In the future, we hope to perform additional analysis, such as electron microscopy, on our samples, so that we can directly correlate these albedo changes with specific mineralogical or chemical changes that may occur in the irradiated layer.

Unlike the darkening trend, we observe an increase in 3-µm absorption band depth after laser irradiation. We speculate that this increase in band depth is related to a structural change induced by the laser, rather than indicative of the formation of newly hydrated species. This is supported by the observation that other spectral features in this wavelength range, such as those due to M-OH near 2.3 µm (Clark et al. 1990, Cloutis et al. 2011) and organics near 3.4 µm (Clark et al. 2009) also appear to increase slightly with irradiation (Figure 2). Perhaps similarly, we have also observed that the band depth is sensitive to packing or intergrain porosity of our pellet samples, as the band depth of the 2.72 µm band was significantly greater in loose CM simulant powder compared with the pellets irradiated here (~0.44 vs. 0.27). While we don't have the ability to irradiate loose powders in our current setup, we hope to modify our system to be able to conduct these types of experiments in the future.

Interestingly, if the increase with band depth with laser irradiation is indeed due to some structural modification of the sample via laser irradiation, then it is not immediately clear why a similar change is not observed after laser irradiation of Murchison pellets (Matsuoka et al. 2015, 2020). We cannot attribute this to a difference in laser fluence, as those single shot experiments used fluences between $0.06 - 1.2$ J / cm$^{-2}$ and found that the band depth decreased in all cases. It seems likely that the different trend observed for the band depth suggests that space weathering effects induced by laser irradiation may be strongly dependent on composition of the hydrated minerals present, as from the perspective of the hydrated minerals, our samples are likely more representative of CI chondrites (see Section 4.2). We also note that the spectra reported in Matsuoka et al.

(2015, 2020) also appear to have a distinct absorption band centered near 3.0 µm, which appears more prominent than what is seen in a spectra of Murchison taken in a vacuum environment (Beck et al. 2010, Takir et al. 2019). If this difference is not a consequence of Murchison's heterogeneity (see above), then it suggests that the laser irradiated samples contained some terrestrial $H_2O$ (see Figure 1). The possible presence of terrestrial $H_2O$ and its subsequent removal via laser irradiation was also noted in the study on Murchison chip samples (Thompson et al. 2020). More specifics on whether and how the presence of a significant amount of terrestrial $H_2O$ could affect observed spectral changes induced by space weathering are currently lacking but could be tested in future studies.

*4.3.2 Previous Ion irradiation experiments*

Although studies on silicates such as olivine have shown general agreement between changes in near-infrared reflectance spectra induced by ion irradiation and pulsed laser irradiation (Loeffler et al. 2016), one should not expect, a priori, that this will be the case in other spectral regions. Thus, we also compare our experiments to those performed with ion irradiation.

As mentioned above, Lantz et al. (2015) irradiation Murchison with both $He^+$ and $Ar^+$ and analyzed with sample before and after irradiation with ex situ reflectance spectroscopy. While the spectral changes were noted to be complicated by the presence of terrestrial $H_2O$, they suggested that there was a possible red shift of the 3-µm absorption band as a result of irradiation. Again, using ex situ reflectance spectroscopy, Lantz et al. (2017) irradiated carbonaceous chondrites Mighei, Alais, and Tagish Lake (CM, CI, and ungrouped chondrites, respectively) with $He^+$ ions. These meteorite samples contained the sharp phyllosilicate feature at 2.7 µm, as well as a broader absorption band centered near 2.8 µm. The former, which is indicative of the most aqueously altered phyllosilicates, significantly decreased during irradiation and the band minimum appeared to red shift by between 0.06 and 0.09 µm depending on the sample. Recently, Rubino et al. (2020) irradiated two different serpentine samples with $He^+$ ions but performed spectral analysis in situ. The only notable change in the phyllosilicate feature was a slight red shift in the band minimum by ~0.015 – 0.02 µm, as there were no observable changes in the band shape or intensity (band depth) as a result of irradiation. Most recently, Nakauchi et al. (2021) irradiated saponite and serpentine with $H_2^+$ ions, performing analysis in ambient conditions. They found that the absorption strength of the hydrated features near 3-µm increased slightly with irradiation for both samples but no spectral shift in band minima was noted. The increase in band depth with irradiation was speculated to be due to implanted hydrogen reacting and forming new bonds within the mineral surface, as has been seen previously in anhydrous minerals (see Section 1).

The overall trend observed in these ion irradiation studies suggests that irradiation with heavier ions (He$^+$, Ar$^+$, etc.) may cause the band minimum of the phyllosilicate feature to shift slightly to longer wavelengths, unlike what we observed after laser irradiation. In addition, our results on the increase of the band depth of the phyllosilicate feature with laser irradiation compared with the nominal alteration of this feature's intensity with ion irradiation (Rubino et al. 2020) is not so unexpected, as the laser may be much more efficient in altering the surface structure of these minerals, as it will penetrate more deeply than the ions and melt large regions of the mineral surface. However, comparing these two results with those of Lantz et al. (2017) also suggests that the 2.7 µm feature may be less stable when a range of aqueously altered minerals are present. If this latter trend can be shown in future laboratory studies, then it would provide a better understanding of what mineralogies are needed to properly simulate a specific extraterrestrial environment in the laboratory, as well as help guide the degree to which future observational studies would need to consider space weathering processing in their attempts to determine an asteroid's surface composition.

*4.4 Implications for Space Weathering on Airless Bodies*

Access to the entire 3-µm spectral region both through current asteroid missions, as well as in the near future with the JWST, will greatly increase our understanding of the surface composition of hydrated asteroids. The ability to evaluate the degree to which space weathering may complicate any remote sensing analysis is critically important. Below we discuss the possible role that space weathering may play in altering the mineral reflectance spectra near 3-µm, using our new laboratory results, as well other recent ones.

Rubino et al. (2020) observed that the band intensity of the sharp phyllosilicate feature will not be significantly altered by He$^+$ ion irradiation of Mg-rich serpentines, while we observed that the intensity (i.e. band depth) increases with laser irradiation. Given that we do not observe any significant change in the band shape with irradiation and a number of other factors could also alter the band depth, it would be difficult to uniquely link any band depth variations across an asteroid's surface, such as what has been recently reported for the phyllosilicate feature detected on Bennu (Simon et al. 2020), directly to space weathering. Furthermore, the location of the band minimum will either shift minimally (Rubino et al. 2020) or not at all – as we observe here – as a result of space weathering processes. It is unknown whether this discrepancy in band position is due to the weathering agent (ion or pulsed laser irradiation) or differences in sample mineralogy, but this could be tested in future studies. These two results together suggest that, at least for hydrated asteroids that have undergone extensive aqueous alteration, this sharp phyllosilicate feature may not be strongly affected by space weathering processing and thus could be very useful to determine the composition of hydrated asteroids even if the surface had been subject to a significant amount of space weathering.

For surfaces of carbonaceous asteroids that contain a range of aqueously altered minerals, which may be most prevalent based on spectral studies of a range of carbonaceous meteorite samples (Beck et al. 2010, Takir et al. 2013, Takir et al. 2019, Hiroi et al. 2021), establishing how weathering may affect the 3-μm absorption band may be more difficult. Recent studies have shown that there may be significant change in this spectral region as a result of ion bombardment; this may include a shift in the band minima to longer wavelengths in addition to the attenuation of the sharp phyllosilicate feature (Lantz et al., 2015; Lantz et al., 2017). Determining whether this trend is generally applicable to surfaces with less or mixed degrees of aqueous alteration is something that should be investigated in more detail using experiments that can be performed entirely in situ to minimize contribution from environmental factors, such as adsorbed $H_2O$.

## 5. Conclusions

Here we studied space weathering effects on the 3-μm absorption region in CI and CM simulant pellets induced by pulsed laser in an effort to simulate micrometeorite impacts. Judging from our general analysis at short spectral wavelengths (1.8 μm), we find that the laser irradiation causes the blue spectral slope to decrease until it is relatively flat. Additionally, the sample also darkens initially with laser irradiation but subsequently brightens to about half of its original level after 5 laser pulses. These effects will be able to be studied in greater detail when we perform similar experiments in the visible and near-infrared, where slope changes and changes in albedo will likely be more pronounced.

Focusing on the 3-μm absorption band, we find that laser irradiation causes the band depth to increase by as much as 30%, which we speculate is related to a structural change induced by the laser that increases the effective path length of the infrared light in our sample. Regardless, even with this change in band depth, we find that the shape of the entire absorption band does not change and the band minima of the 2.72 μm phyllosilicate feature shifts less than 0.001 μm after laser irradiation. Together, these findings suggest that this spectral region could be very useful to determine the asteroid composition on surfaces on hydrated asteroids that have undergone extensive aqueous alteration even if the surface had been subject to a significant amount of space weathering. Whether the same conclusion will apply to surfaces containing less aqueously altered, hydrated minerals or a broader distribution of aqueous alteration is currently unclear but could be systematically tested with laboratory experiments performed entirely in situ, which would minimize factors that complicate analysis, such as the presence and contribution from terrestrial $H_2O$.

**Acknowledgements**

BSP was supported by Northern Arizona University's Hooper Undergraduate Research Award.  After publication data will be placed in Northern Arizona University's long-term repository (https://openknowledge.nau.edu).  Currently, the data can be found on our laboratory website (http://www.physics.nau.edu/~loeffler/Submitted-data.html).

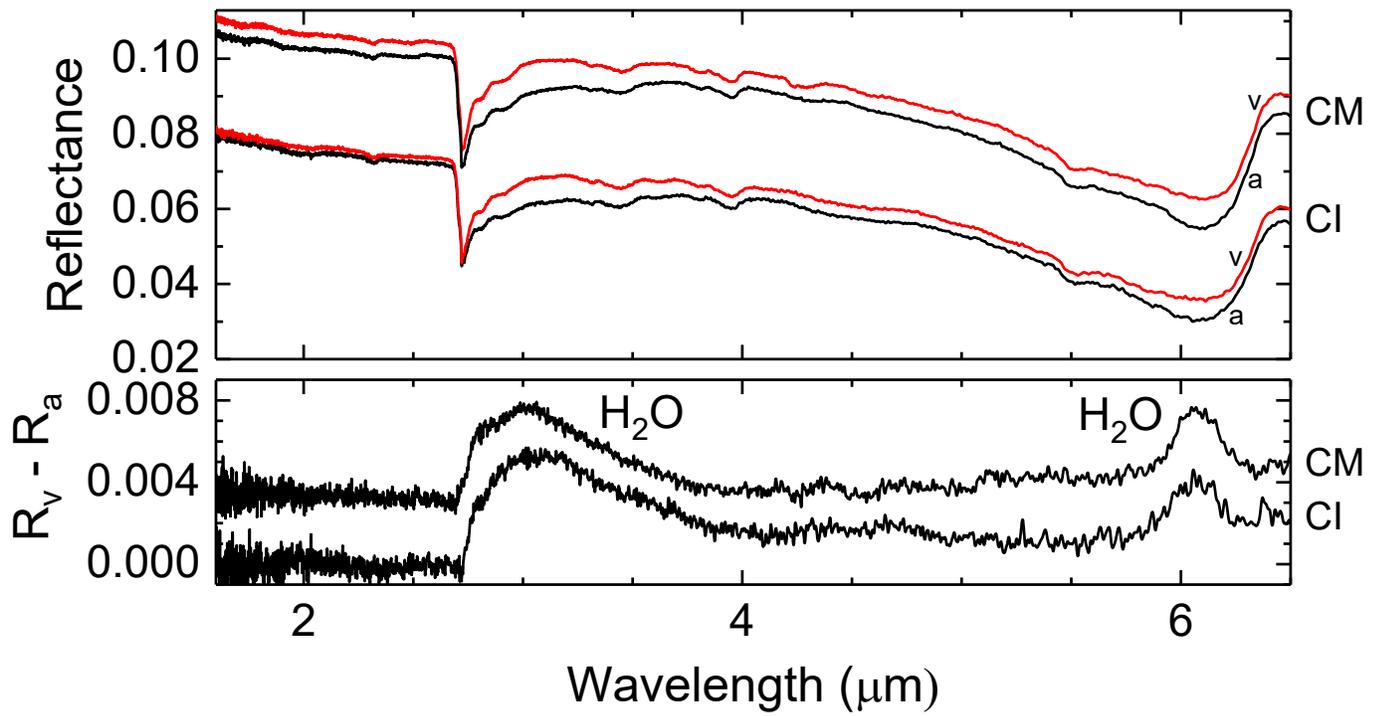

Figure 1. Top: Reflectance spectrum of the CM and CI simulant (<45 μm) pellet under ambient (labeled 'a') and $10^{-8}$ Torr (labeled 'v') conditions. Bottom: Difference spectra of the vacuum spectra ($R_v$) and ambient spectra ($R_a$) for the CM (top) and CI (bottom) pellets.

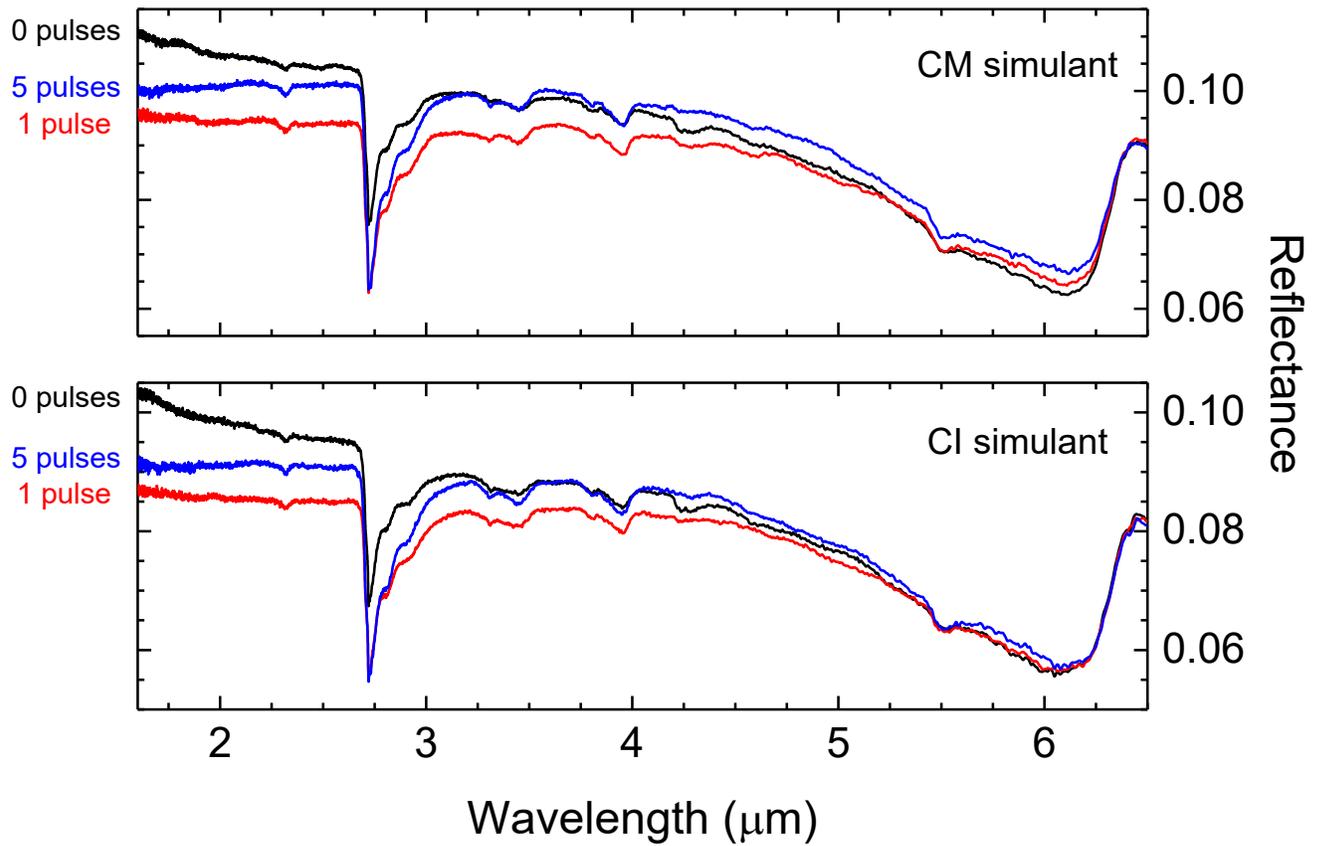

Figure 2. Reflectance spectrum of the CM (top) and CI (bottom) simulant (<45 µm) pellet during pulsed laser irradiation.  In both panels, spectra (from top to bottom at 1.75 µm) are 0, 5, and 1 pulses.

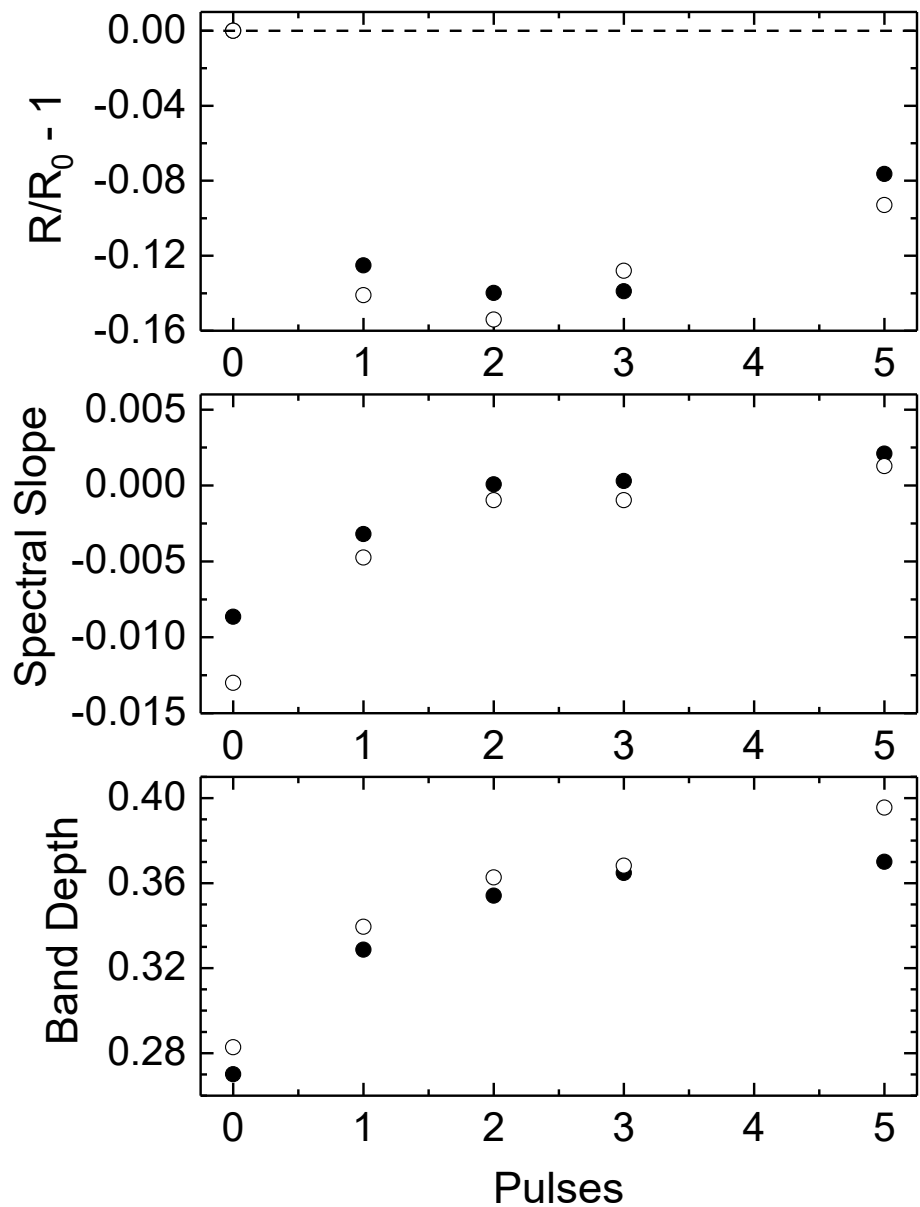

Figure 3. Changes in normalized reflected intensity evaluated at 1.8 μm (top), the spectral slope evaluated at 1.8 μm (middle), and band depth of the 2.72-μm absorption band minimum (bottom) and for the CM (●) and CI (○) simulant (<45 μm) pellets shown in Figure 2 during pulsed laser irradiation.  The dashed line in the top panel indicate the initial normalized reflected intensity of the pellet at 1.8 μm.

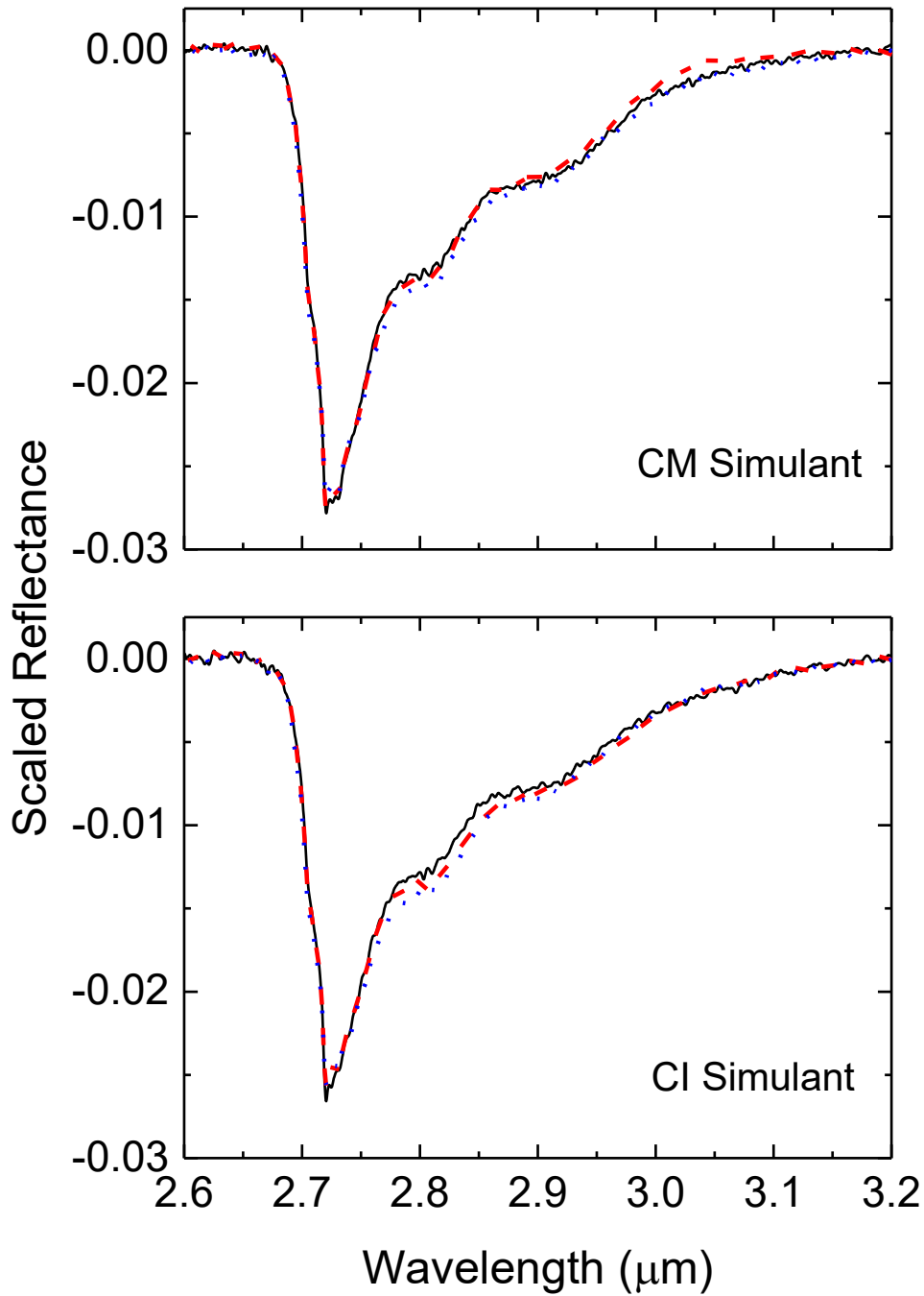

Figure 4. Scaled reflectance spectra of the 3-μm absorption region for the CM (top) and CI (bottom) simulant (<45 μm) pellets after 0 (black, solid line), 1 (red, dashed line) and 5 (blue, dotted line) laser pulses. The spectra were produced by removing a linear continuum and multiplying each spectrum by the following: 0 pulses (no correction), 1 pulse (CM, 0.90; CI, 0.89), 5 pulses (CM, 0.72; CI, 0.72).